# Polaron population controlled of modulation wave vector Manganites

P. R. Sagdeo

UGC-DAE Consortium for Scientific Research, University campus, Khandwa road Indore, India- 452017.

And

Bhabha Atomic Research Centre, Visakhapatnam Project, Autonagar Campus Visakhapatnam, India-530012.

The growth of charge-ordering phenomenon in calcium doped lanthanum manganites $La_{1-x}Ca_xMnO_3$; x = 0.62, 0.67 has been studied using transmission electron microscopy (TEM). The diffuse intensity characteristic of short range charge ordering is observed even at room temperature. This diffuse intensity has the local intensity maxima at the incommensurate positions ±q. With decreasing temperature, the intensity 'I' of super-lattice charge-ordered spot and the magnitude of its modulation wave vector 'q' increases and there exists a linear relationship between 'I' and 'q' of super-lattice charge ordered spot. This linear dependence can be explained taking in to account the population and the columbic interactions between the correlated polarons. It appears that the columbic interactions between the correlated polarons lead to the coherent distributions of $e_g$ charge carriers and thereby giving rise to the uniformly periodic superlattice charge-ordered state to these manganites.

## Introduction:

The phase competition in colossal magneto-resistive (CMR) manganites is one of the hottest topics in condensed matter physics. These systems exhibit varieties of structural electronic and magnetic phase transition, and complex phase coexistence over wide temperature and composition range[1-4]. In mixed valence manganites such as $La_{1-x}Ca_xMn^{+3}_{1-x}Mn^{+4}_xO_{3;}$ the $Mn^{+3}$ and $Mn^{+4}$ ions are randomly distributed throughout the solid, but under certain temperature and composition range; these charges redistribute themselves coherently, this state is called *charge-ordered* state. Even though vast literatures are available on these charge ordered systems[5,6], but clear understanding regarding the origin and the nature of the charge order state in manganites is still unclear. The crucial role of anti-ferromagnetic spin alignment as a origin of charge ordering has been discussed in some literatures[7,8], on the other hand the presence of strong columbic interactions is also thought to be the main mechanism for charge ordering in these manganites[9,10]. Along with the origin of charge ordering in manganites, the issue of temperature dependence of modulation wave vector of the charge order state also matter of debate. The pictures like “charge density wave” and “discrete nature of charge ordering due to localization of $e_g$ charge carriers at the periodic site” have been sketch to explain the temperature dependence of modulation wave vector in these manganites[11,12]. In order to clarify the above issue, we have very carefully studied the phenomenon of charge ordering in $La_{1-x}Ca_xMnO_3$; x = 0.62, 0.67. Our results suggest that, the columbic interaction between the correlated polarons leads to the Wigner crystallization and thus gives rise to uniform periodic super-lattice (commensurate or incommensurate) charge ordered state to these manganites. Further our results suggests that, with decrease in temperature the population of these correlated polarons increases which, controls the magnitude of modulation wave vector q in these manganites.

**Experimental:**

The polycrystalline samples of $La_{1-x}Ca_xMnO_3$ (x=0.62, 0.67) were prepared by standard solid-state reaction route[13,14]. The structural characterizations of the as-prepared samples were carried out using powder x-ray diffraction[13] (XRD) and 200kV FEI-TECNAI-20-G[2] transmission electron microscope (TEM). The TEM is equipped with Gatan double tilt low background liquid nitrogen holder and a CCD camera (Mega-View SIS). The sample for TEM experiments was prepared by the conventional method as described elsewhere[14], the magnetization measurements were performed on Vibrating Sample Magnetometer (VSM) and the resistivity measurements were carried out on self developed four-probe resistivity measurement system [15].

**Results and discussions:**

Figure1 (a) shows the representative room temperature powder XRD pattern for the studied samples, where as figure 1(b) is the selected area electron diffraction pattern (SAD) for the same sample along [010]-zone axis. The XRD patterns of the as prepared samples were indexed using orthorhombic structure with the space group *Pnma*[13]. The absence of any unaccounted peak confirms the structural phase purity of these samples. The presence of diffuse electron intensity (as indicated by arrows) characteristics of room temperature charge ordering[14] can be seen near the fundamental Bragg spots. These diffuse scatterings are similar to those reported earlier using x-ray and neutron scattering studies[16-23]. We have recorded such electron diffraction patterns in the temperature interval of 300 K to 106 K. A representative sequence of SAD pattern for $La_{0.38}Ca_{0.62}MnO_3$ recorded at various temperatures is shown in figure2. The similar results have been obtained for the other studied samples[14]. From the figure 2 it is clear that with the decrease in the temperature the intensity 'I' and the magnitude of modulation wave vector 'q' increases and gives rise to the strong super-lattice charge order spot. Figure3 shows the intensity profile at room temperature (along the *white line* shown in figure 1-b) along a*. From the intensity profile shown in the Figure3 it is clear that the diffuse scattering has the local intensity maxima at the incommensurate positions ±q. A representative variation of modulation vector 'q' as function of temperature for $La_{0.38}Ca_{0.62}MnO_3$ is shown in figure4 (a), where as Figure4 (b) and (c) shows the temperature dependence of intensity and the peak width (FWHM) for the same super-lattice peak. These results are consistent with those reported earlier by similar electron diffraction studies[24,25]. Figure-5 show the high resolution lattice fringes due to charge ordering at various temperatures, from the figure-5 it is clear that with decrease in temperature the coherency of the charge order state increases, this result is in consistence with reported by Zuo eta al[24]. Thus with figure-5 we can understand the increase in the intensity and decrease in the FWHM of super-lattice charge ordered peak as shown in the figure-4 (b) and (c). The important point here is to note that the variation of modulation wave vector 'q' cannot be understood with the present high resolution images. Millard et al[11] have proposed the charge density wave picture to explain the temperature variation of 'q', where as Chen et al[12] proposed that the discrete nature of charge ordering due to localization of $e_g$ charge carriers at the periodic site is responsible for discontinuous change in the modulation across transition. For Manganites it should be noted that the nature of charge-ordered state is a matter of debate, the charge density wave model, Wigner crystallization model and bi-stripes model are reported in the literatures[11,12],[26-30].

It is interesting to note that the signature of the charge ordering is present even at room temperature, where the sample is in the paramagnetic state. Fig. 6(a) and 6(b) are the resistivity and magnetization for the studied samples, as clear from the magnetization studies that both the samples are in the paramagnetic state at the room temperature, where as the TEM studies clearly reviles the presence of super-lattice charge order peaks even at this temperature, hence it seems that the anti-ferromagnetic spin alignment is not necessary condition for the charge ordering in manganites. It should be noted that the temperature at which the slope of the magnetization reverses its sign; at the same temperature the intensity (I), modulation wave vector (q) and the peak width (FWHM) almost (nearly) saturates. Thus it appears that the even though the anti-ferromagnetic alignment is not necessary for occurrence of charge ordering in these manganites but these anti-ferromagnetic interactions between the Mn ions helps to stabilizes the charge order state in these manganites.

In this section we present our key results. From the figure-3 it is clear that the diffuse intensity has the local intensity maximum at the in-commensurate positions $\pm$ q. For manganites such diffuse has already been observed using resonant x-ray and neutron scattering experiments[16-23]. Using these studies it has been well established that such diffuse intensity is due to the polarons associated with the Jahn-Teller distortion[16,20]. It should be noted that these polarons are formed due to the localization of $e_g$ electrons at the Mn site; which leads to the distortion of $Mn\text{-}O_6$ due to the Jahn-Teller effect[16,20]. With decrease in temperature the intensity of the charge-order peak grows and gives rise to sharp charge order peak. This increase in the intensity of the charge ordered spot has been attributed to the increase in the population of correlated Jahn-Teller polarons due to localization of $e_g$ electrons at the Mn site[17] which further leads to the increase in the coherency of the charge order state in these manganites[20,22] and that what has been observed in the present studies. Thus with decrease in the temperature the population and the coherency of Jahn-Teller polarons increases. The important point here is to note is that there exist one-to-one correspondence between the magnitude of wave vector 'q' and Intensity 'I' of super-lattice charge order spot (Figure-4) i.e. with decrease in the temperature the intensity of diffuse super-lattice spots and the magnitude of its modulation wave vector increases.

From the above discussions it is clear that *"with decrease in temperature the number of polarons associated with the Jahn-Teller distortion increases"* [17,20,22]. It should be noted that the Hall measurements studies also supports the localization of charge carriers with decreasing temperature for these charge order manganitges[31] and the localization of charge carriers at Mn site leads to formation of Jahn-Teller polarons. In this paper we propose that it is the numbers of these polarons associated with the Jahn-Teller distortion and the columbic interaction between them controls the magnitude of modulation wave vector 'q'. This can be understood as discussed below.

Consider a *real space* situation as shown in the figure-7 (it should be noted that the wave vector 'q' have the dimension of reciprocal space), consider a line joining point A and B as shown in the Figure-7, if only *one* electron is allowed to localize on this line; then this *single* electron can occupy any position on line AB. Now if one more electron is *added* on the same line AB then in order to have minimum columbic repulsive energy, these *two* electrons will get localized at the end points on the line AB i.e. one electron will get localized at point A and other will be localized at point B. From the figure 6 it is clear that if these additions of electrons continue then in order to minimize the columbic potential energy the periodicity of $e_g$ electron will get modified. Thus from the figure 6 it is clear that the magnitude of wave vector is effectively controlled (or dictated) by the number of localized charge carriers, this phenomenon is called Wigner Crystallization and it has been well supported for charge order manganites through the different techniques[26,27]. Thus with the increase in the number of electrons on line-AB, leads to the decrease in the distance between the two electrons in real space, hence in the reciprocal space the magnitude of modulation wave vector shows the opposite trend, i.e. the modulation wave vector increases with increase in the numbers of polarons associated with Jahn-Teller distortion. If the above argument is true for the charge ordered manganites then there must exist a linear relationship between the intensity and the magnitude of modulation wave vector 'q'. In order to verify this we have plotted the magnitude of modulation wave vector 'q' on Y axis and its intensity 'I' on X axis (figure-8). As clear from the figure-8 that, there exists a linear relationship between the populations of localized polarons (the intensity) and wave vector 'q'. It is important to note that the similar linear relationship can also be extracted from the various data reported earlier, such as by *Shimomura et al*[22]. This further confirms that the *magnitude of wave vector q is controlled by the numbers of the polarons associated with Jahn-Teller distortion.* It should be noted that the strong electron lattice coupling due to the Jahn-Teller effect demands that the $e_g$ electron of $Mn^{+3}$ should get localized at the Mn site[12,30], but in order to minimize the coulomb repulsive energy these itinerant $e_g$ electrons will try to move as away as possible. Thus it may be possible that the itinerant $e_g$ electron is close to the Mn site but not exactly at the Mn site[32]. Thus due to the presence of intrinsic columbic repulsion the periodicity of $e_g$ electron (i.e. magnitude of modulation wave vector 'q') gets modified and in doing so it may not be exactly at Mn site but may along the 'a' axis of *Pnma*. This may be origin of proposed weakening of electron lattice coupling in these manganites[28]. In the present scenario it is difficult to understand the localization of $e_g$ charge carriers in between two Mn atoms i.e. along the 'a' axis of *Pnma*. Such localizations suggest that the $e_g$ electron have weak lattice coupling[28,29] and the $e_g$ electrons may have the localized and broad wave characters[33]. A very careful high-resolution synchrotron x-ray photoelectron spectroscopy studies may be useful in this regard. At this stage we would like to bring in notice that Sánchez *et al*[34] have observed the lack of jahn-teller distortion (and there by weakening of electron lattice coupling) in Ga substituted $LaMnO_3$ and attributed this to the dilution of $Mn^{+3}$ ions in the sample. The present samples $La_{0.33}Ca_{0.67}Mn^{+3}_{0.33}Mn^{+4}_{0.67}O_3$ and $La_{0.38}Ca_{0.62}Mn^{+3}_{0.38}Mn^{+4}_{0.62}O_3$ certainly have lower concentration of $Mn^{+3}$ as compared to that of $Mn^{+4}$ this may also lead to the further weakening of electron lattice coupling as suggested by Sánchez et al[34]. It is interesting to note that various electron microscopy[28,29], optical[35,36], electrical[37] thermal[38] and theoretical[l1,39,40] studies on manganites suggests that the charge-ordered state of manganites is a Charge Density Wave (CDW) or Peierls-instability of $e_g$ electrons. At this juncture we would like to emphasize that the phenomena of charge density wave for manganites theoretically treated using Ginzburg-Landau theory. Using the same Ginzburg-Landau theory W.L Mcmillan[41,42] shown that for in-commensurate structure the intensity scales with wave vector ‘q’ (Intensity can be characterized as a order parameter in disorder to order structural phase transition). But this scaling is dose not appears to be linear. This issue needs attention from theoreticians and experimentalists and need to be address in greater details. It should be noted that the quantum phase transition to charge-ordered and Wigner-crystal states under the interplay of lattice commensurability and long-range coulomb interactions has been discussed in the literature and theoretically it has been pointed out that the electron density also plays a crucial role in such transition[43]. For manganites the lattice commensurability changes with hole doping hence the possibility of quantum phase transition from charge ordered state to Wigner crystal state cannot be ruled out as s function of temperature and doping. From the experiments followed by the analysis presented in this paper it appears that it is number of localized charge carriers and the long range columbic interactions between them controls the commensurability of charge order state in manganites. For charge density wave the possibility of *columbic* interactions have been neglected in the Peierls original book, but many theoreticians[44-47] realized that, it is necessary to take in the account the effect of nearest neighbor on-site Coulomb interactions to explain the experimental facts. Hence the exact nature of the charge order state of manganites may not be universal hence still remains the open question, but it appears that its long range columbic interactions which stabilizes the super-lattice (commensurate or incommensurate) so called charge order state in these manganites.

**Conclusions:**

From the electron diffraction experiments at various temperatures we have shown that the short-range polaron correlation via electron-electron interactions is present even at room temperature. With decrease in temperature this polaron correlation grows as the coherent superlattice charge order state. It appears that the the anti-ferromagnetic alignment is not necessary for occurrence of charge ordering in these manganites but these anti-ferromagnetic interactions between the Mn ions helps to stabilizes the charge order state in these manganites. The temperature dependence of modulation wave vector can be explain by considering the population of localized Jahn-Teller polarons and the columbic interactions between them.

**Acknowledgements:** Author sincerely acknowledges Dr. N.P. Lalla for his valuable guidance. Author is also thankful to Prof. T.V. Ramakrishanan, and Prof. D. Pandey for the valuable discussions during SERC School held at BHU. Authors sincerely thank Prof. Ajay Gupta, Dr. P Chaddah for their encouragement and P. B Littlewood for valuable email discussions and for supplying important references. Aurhor also acknowledge Department of Science and Technology Government of India and UGC-DAE-CSR Indore for research fellowship.

**References:**

1. P.R. Sagdeo and N. P. Lalla Phys Rev B **78**, 174106 (2008).
2. M. Fath, S. Freisen, A. Menovsky, Y. tomioka, J. Aarts and J. A. Mydosh, Science **285** 1540 (1999).
3. P.R. Sagdeo, Shahid Anwar, N.P. Lalla, Phys Rev B **74**, 214118 (2006).
4. D. D. Sarma, Dinesh Topwal, U. Manju, S. R. Krishnakumar, M. Bertolo, S. La Rosa, G. Cautero, T. Y. Koo, P. A. Sharma, and S.-W. Cheong, Phys Rev Lett, **93**, 097202 (2004).
5. Rao, C. N. R and Raveau, B. Colossal Magnetoresistance Charge Ordering and Related Properties of Manganese Oxides, World Scientific (1998).
6. Tokura, Y. Colossal Magneto-resistive Oxides, Gordon and Breach Science Publisher (2000).
7. I. V. Solovyev and K. Terakura Phys Rev Lett, **83**, 2825 (1999).
8. Z. Popovic and S. Satpathy Phys Rev Lett, **88**, 197201 (2002).
9. S. K. Mishra, Rahul Pandit, and Sashi Satpathy Phys. Rev. B **56**, 2316 (1997).
10. D. P. Kozlenko, L. S. Dubrovinsky, B. N. Savenko, V. I. Voronin, E. A. Kiselev and N. V. Proskurnina Phys Rev B **77,** 104444 (2008).
11. G. C. Milward, M. J. Caldero´n & P. B. Littlewood Nature **433**, 607 (2005).
12. C. H. Chen, S. Mori, and S-W. Cheong **83**, 4792 (1999).
13. P.R. Sagdeo, Shahid Anwar, N.P. Lalla, Powder Diffraction **21**, 40 (2006).
14. P.R. Sagdeo, Shahid Anwar, N.P. Lalla, Solid State Comm. **137,** 158 (2006).
15. P.R. Sagdeo, Shahid Anwar, N.P. Lalla, J Magnetism and Magnetic Materials **306,** 60 (2006).
16. L. Vasiliu-Doloc, S. Rosenkranz, R. Osborn, S. K. Sinha, J. W. Lynn, J. Mesot, O. H. Seeck, G. Preosti, A. J. Fedro, and J. F. Mitchell, Phys. Rev. Lett. **83**, 4393 (1999).
17. C. P. Adams, J. W. Lynn, Y. M. Mukovskii, A. A. Arsenov, and D. A. Shulyatev, Phys. Rev. Lett. **85**, 3954 (2000).
18. Dai, J. A. Fernandez-Baca, N. Wakabayashi, E. W. Plummer, Y. Tomioka, and Y. Tokura, Phys. Rev. Lett. **85**, 2553 (2000).

19. T. J. Sato, J. W. Lynn, and B. Dabrowski, Phys. Rev. Lett. **93**, 267204 (2004).
20. S. Shimomura, N. Wakabayashi, H. Kuwahara, and Y. Tokura, Phys. Rev. Lett. **83**, 4389 (1999).
21. M. v. Zimmerman, J. P. Hill, D. Gibbs, M. Blume, D. Casa, B. Keimer, Y. Murakami, Y. Tomioka, and Y. Tokura, Phys. Rev. Lett. **83**, 4872 (1999)
22. S. Shimomura, T. Tonegawa, K. Tajima, N. Wakabayashi, N. Ikeda, T. Shobu, Y. Noda, Y. Tomioka, and Y. Tokura, Phys. Rev. B **62**, 3875 (2000)
23. V. Kiryukhin, B. G. Kim, T. Katsufuji, J. P. Hill, and S-W. Cheong, Phys. Rev. B **63**, 144406 (2001).
24. J. Tao and J. M. Zuo, Phys. Rev. B **69**, 180404 (2004).
25. A. P. Ramirez, P. Schiffer, S. W. Cheong, C. H. Chen, W. Bao, T. T. M. Palstra, P. L. Gammel, D. J. Bishop, and B. Zegarski, Phys. Rev. Lett. **76**, 3188 (1996).
26. P. G. Radaelli, D. E. Cox, L. Capogna, S.-W. Cheong, and M. Marezio, Phys. Rev. B **59**, 14440 (1999).
27. R. Wang, J. Gui, Y. Zhu, and A. R. Moodenbaugh, Phys. Rev. B **61**, 11946 (2000).
28. J. C. Loudon, S. Cox, A. J. Williams, J. P. Attfield, P. B. Littlewood, P. A. Midgley, and N. D. Mathur, Phys. Rev. Lett. **94**, 097202 (2005).
29. J. C. Loudon, S. Cox, N. D. Mathur and P. A. Midgley, Philosophical Magazine, **85**, 999 (2005).
30. S. Mori, C.H. Chen, and S-W. Cheong, Nature **392**, 473 (1998).
31. I. Gordon, P. Wagner, A. Das, J. Vanacken, V. V. Moshchalkov, Y. Bruynseraede, W. Schuddinck , G. Van Tendeloo, M. Ziese, G. Borghs, Phys. Rev. B **62**, 11633 (2000).
32. V. Ferrari, M. Towler, P.B. Littlewood, Phys. Rev. Lett. **91** 227202 (2003).
33. T. V. Ramakrishnan, H. R. Krishnamurthy, S. R. Hassan, and G. Venketeswara Pai, Phys. Rev. Lett. **92,** 157203 (2004).
34. M. C. Sánchez, J. García, G. Subías, and J. Blasco, Phys. Rev. B 73, 094416 (2006).
35. P. Calvani, G. De Marzi, P. Dore, S. Lupi, P. Maselli, F. D'Amore, S. Gagliardi, and S.-W. Cheong, Phys. Rev. Lett. **81**, 4504 (1998).

36. N. Kida and M. Tonouchi, Phys. Rev. B **66**, 024401 (2002).
37. S. Cox, J. Singelton, R. D. Mcdonald, A. Migliori and P. B. Littlewood, Nature Materials, **7,** 25 (2007).
38. S Cox, J. C. Lashley, E. Rosten, J. Singleton, A. Jwilliams and P.B. Littlewood J. Phys.: Condens. Matter **19,** 192201 (2007).
39. L. Brey, Phys. Rev. Lett. **92**, 127202 (2004).
40. L. Brey and P. B. Littlewood, Phys. Rev. Lett. **95**,117205 (2005).
41. W. L. Mcmillan Phys Rev B **12,** 1187 (1975).
42. W. L. Mcmillan Phys Rev B **14,** 1496 (1976).
43. Y. Noda and M. Imada, Phys. Rev. Lett. **89**, 176803 (2002).
44. J. E. Hirsch, Phys. Rev. Lett. **51**, 295 (1983).
45. S.N. Dixit and S. Mazumdar, Phys. Rev. B **29**, 1824 (1984).
46. B. Horovitz and J. Solyom, Phys. Rev. B **32**, 2681 (1985).
47. Vieri Mastropietro, arXiv:cond-mat/0111164 v1 9 Nov (2001-2006).

## Figure Captions:

**Figure1(a):** A representative room temperature powder x-ray diffraction pattern for $La_{0.38}Ca_{0.62}MnO_3$ refined with the space group *Pnma* absence of any unaccounted peak confirms the single phase nature of the studied sample(s).

**Figure1(b):** Selected area electron diffraction pattern for $La_{0.38}Ca_{0.62}MnO_3$ along [010] zone axis taken at room temperature, the diffuse electron intensity characteristics of room temperature charge ordering can be seen near fundamental Bragg spot and indicated by arrows.

**Figure 2:** Sequence of sections of selected area electron diffraction patterns for $La_{0.38}Ca_{0.62}MnO_3$ collected at various temperatures.

**Figure 3:** Intensity profile at room temperature. Along a* (along the white line shown in figure 1b). The diffuse scattering has the local intensity maxima at the in-commensurate positions ±q.

**Figure 4:** (a) Variation of wave vector q for $La_{0.38}Ca_{0.62}MnO_3$ as a function of temperature. (b) Intensity of super lattice spots as a function of temperature. (c) FWHM of super lattice spot as a function of temperature. A one-to-one correlation between the intensity and modulation of wave vector can be clearly seen from figures (a) & (b).

**Figure 5:** High Resolution lattice fringes due to charge ordering in $La_{0.38}Ca_{0.62}MnO_3$, increase in the structural coherency is evident with decrease in temperature.

**Figure 6:** Temperature dependence of resistivity and magnetization for $La_{0.33}Ca_{0.67}MnO_3$ and $La_{0.38}Ca_{0.62}MnO_3$.

**Figure 7:** One-dimensional Wigner crystallization due to the onsite Coulomb interactions between $e_g$ electrons in charge order manganites, explaining the linear relation between the modulation vector ‘q’ with respect to the polaron population. Thus the population of localized polarons purely controls the magnitude of modulation vector.

**Figure 8:** Linear dependence of modulation wave vector ‘q’ with polaron population suggesting the presence of columbic interaction in charge ordered manganites.

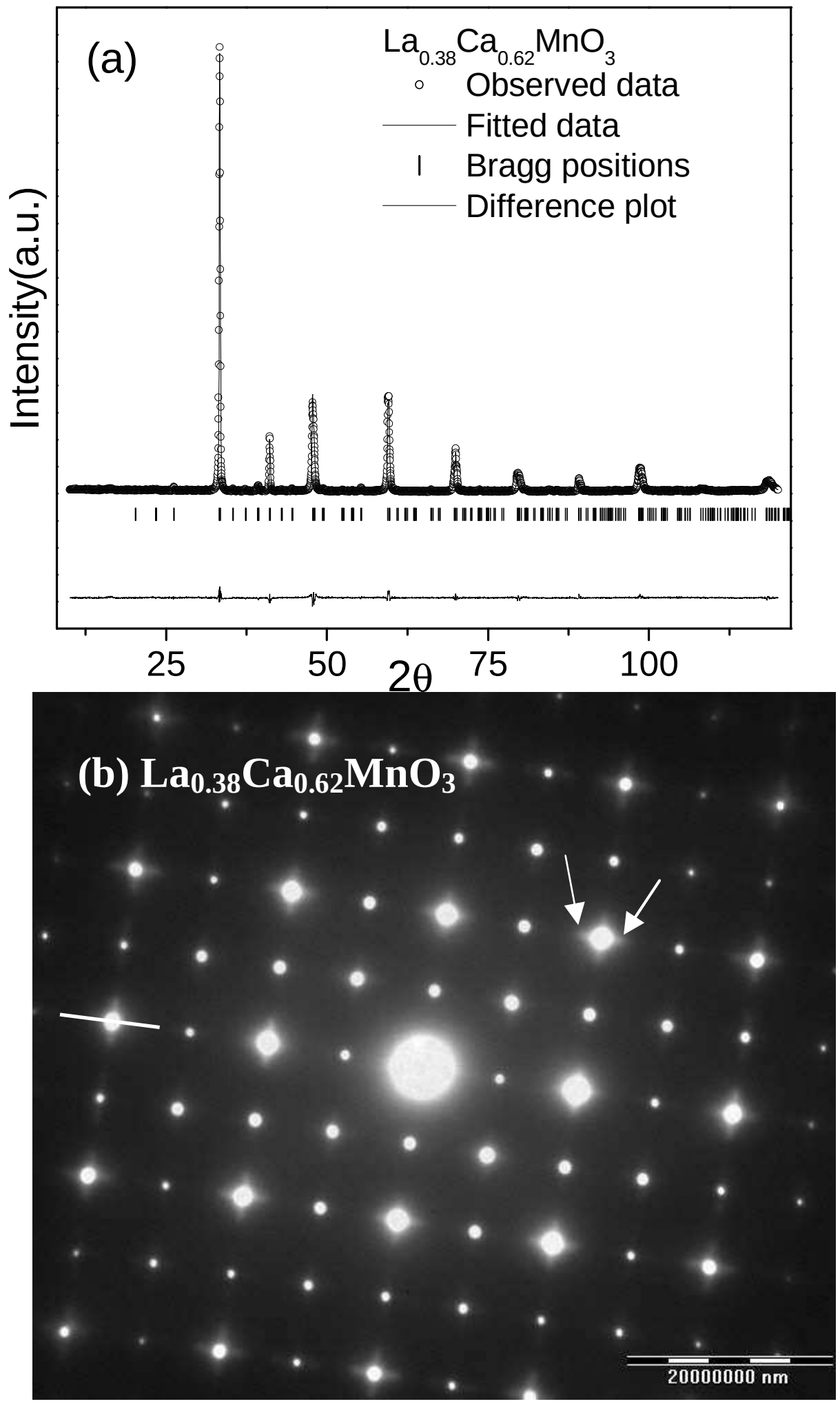

**Fig. 1 (a) and (b)**

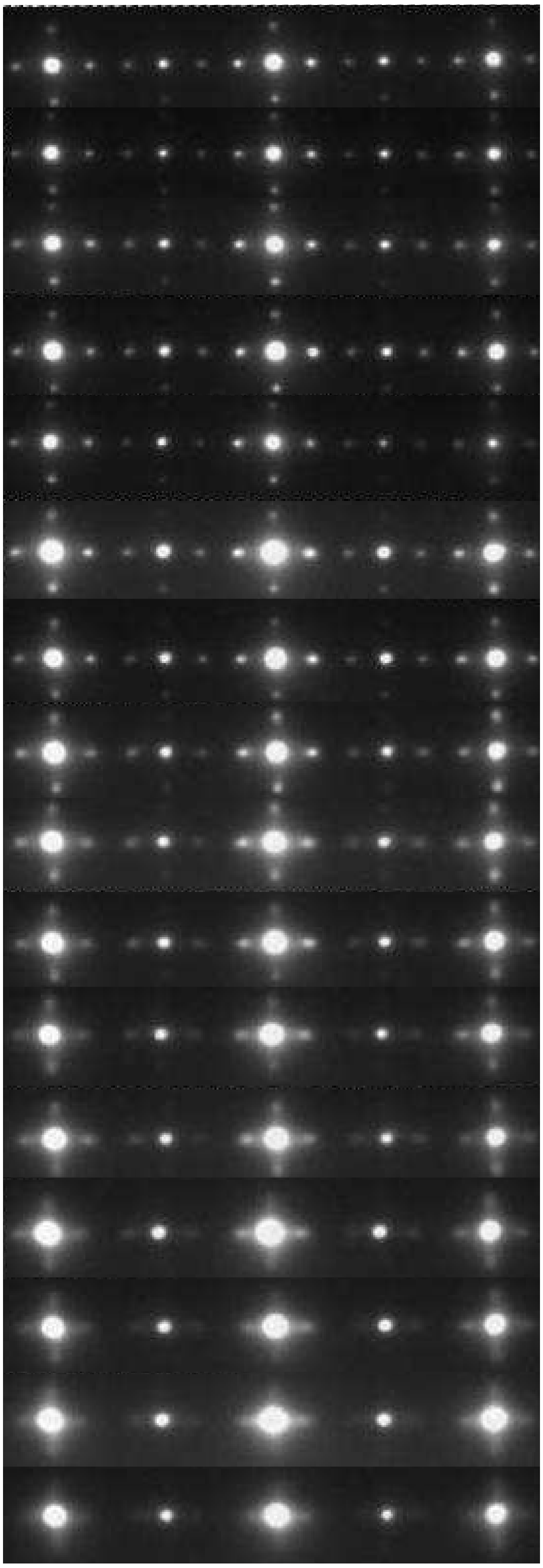

**Fig. 2**

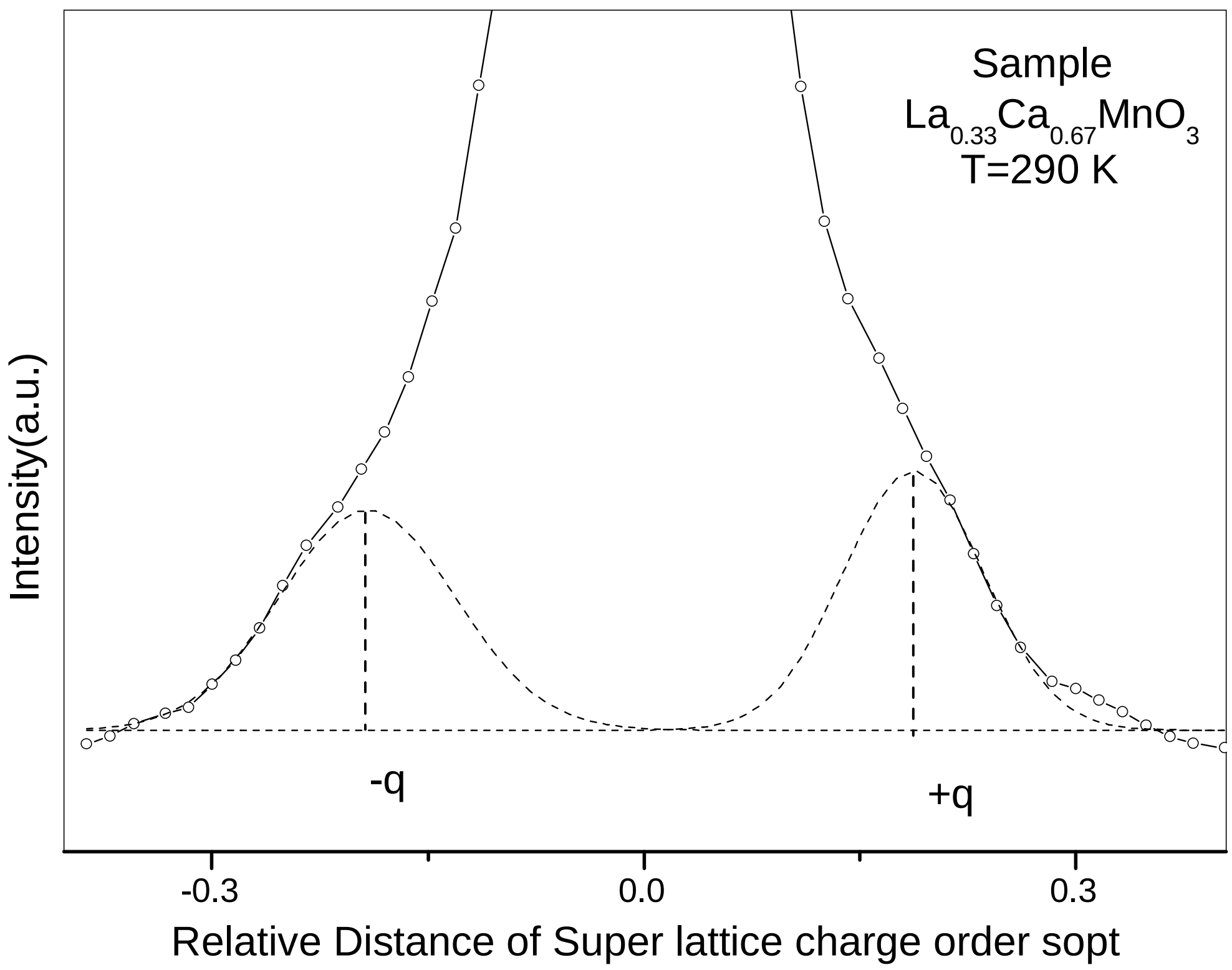
Sample
La0.33Ca0.67MnO3
T=290 K
Intensity(a.u.)
-q
+q
-0.3
0.0
0.3
Relative Distance of Super lattice charge order sopt

Fig. 3

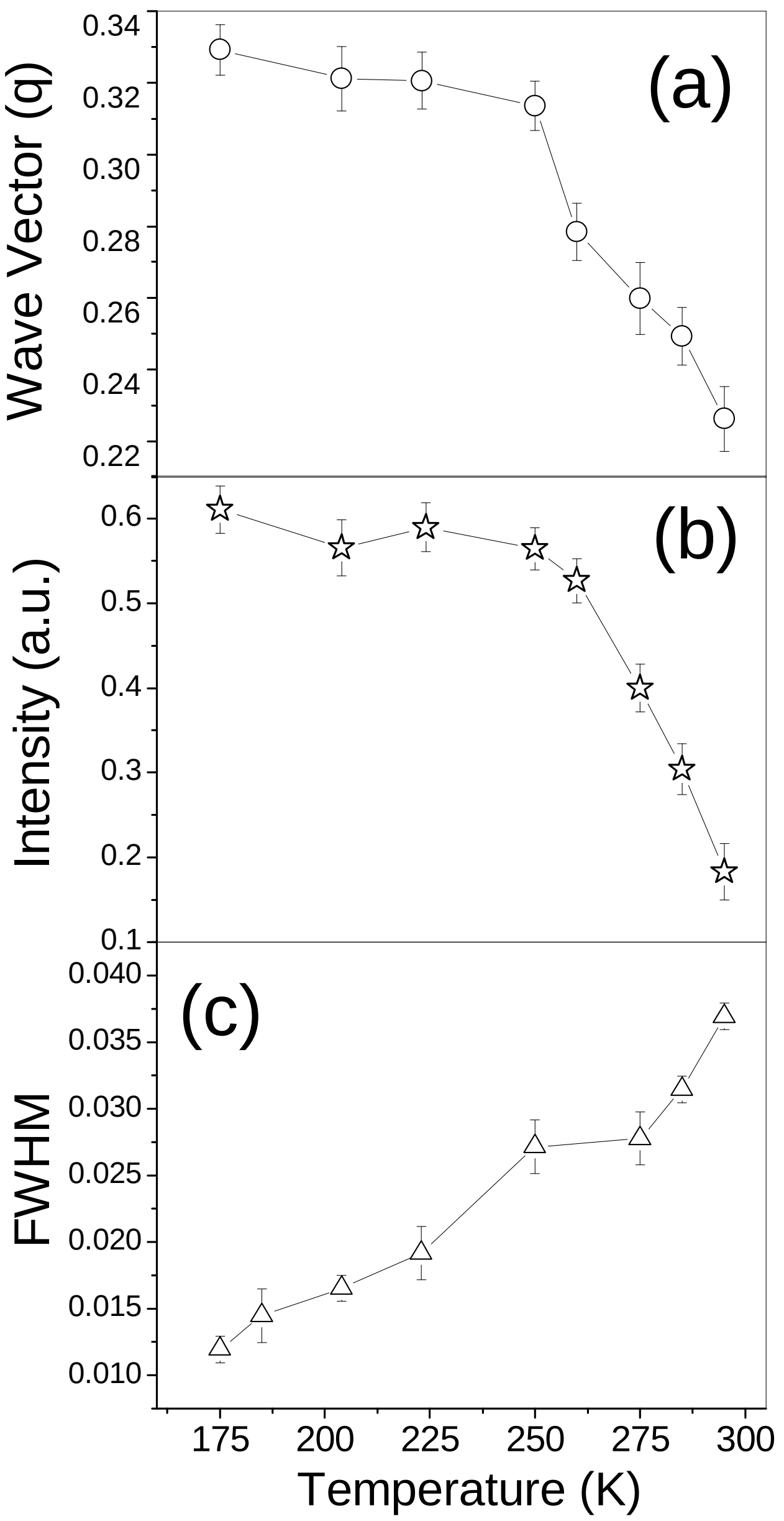

***Fig. 4 (a), (b) and (c)***

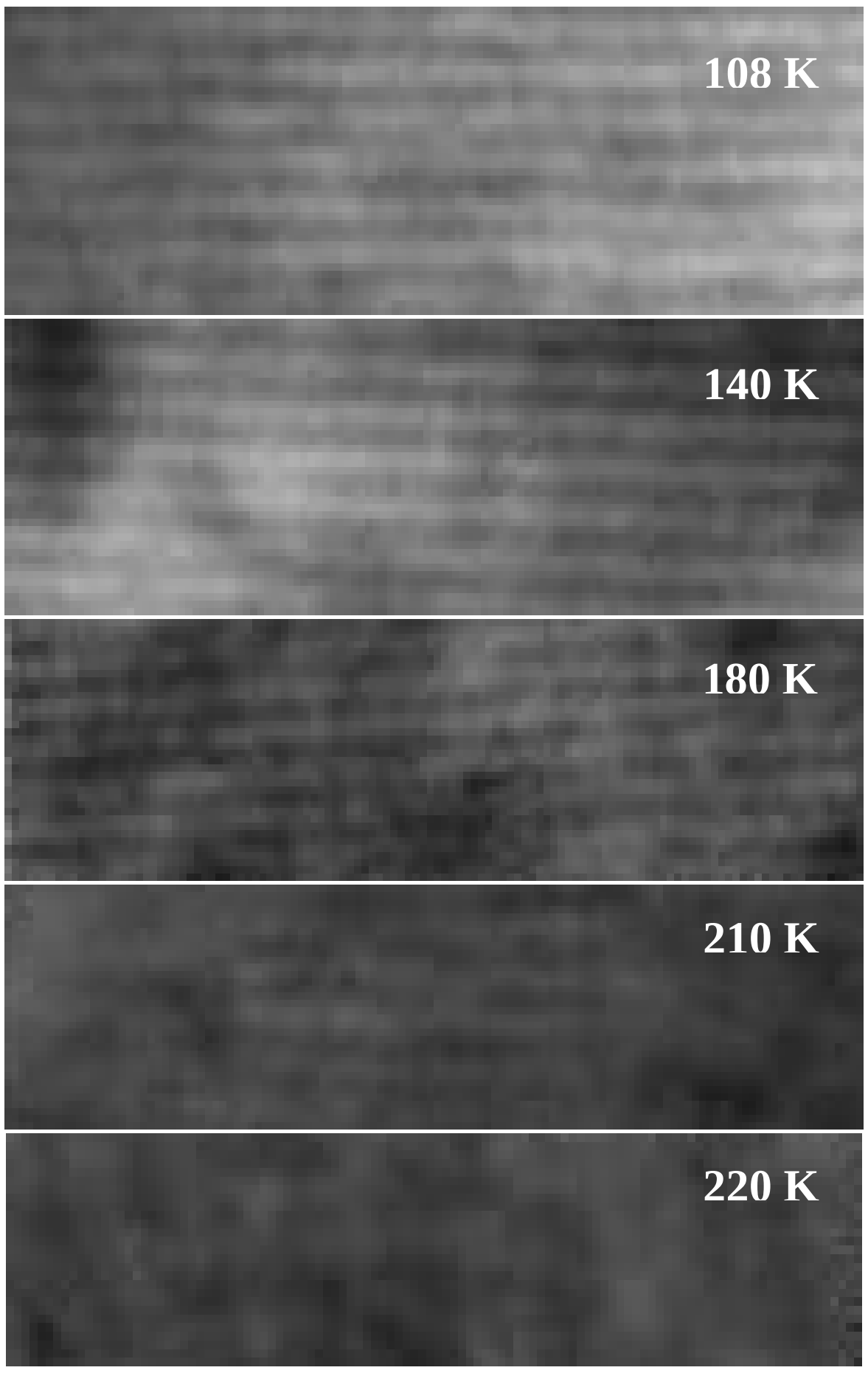

**Fig. 5**

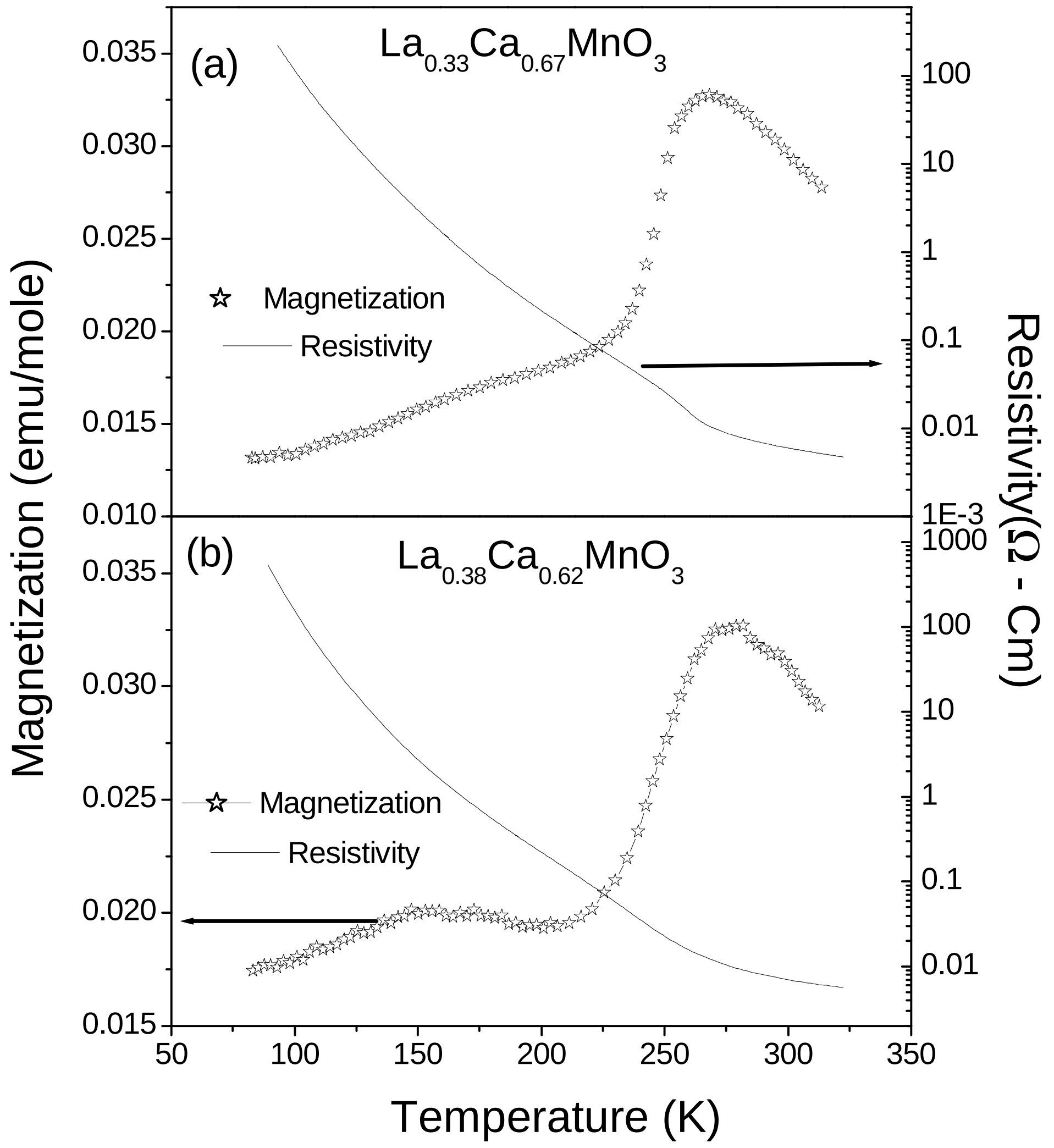

**Fig. 6 (a) and (b)**

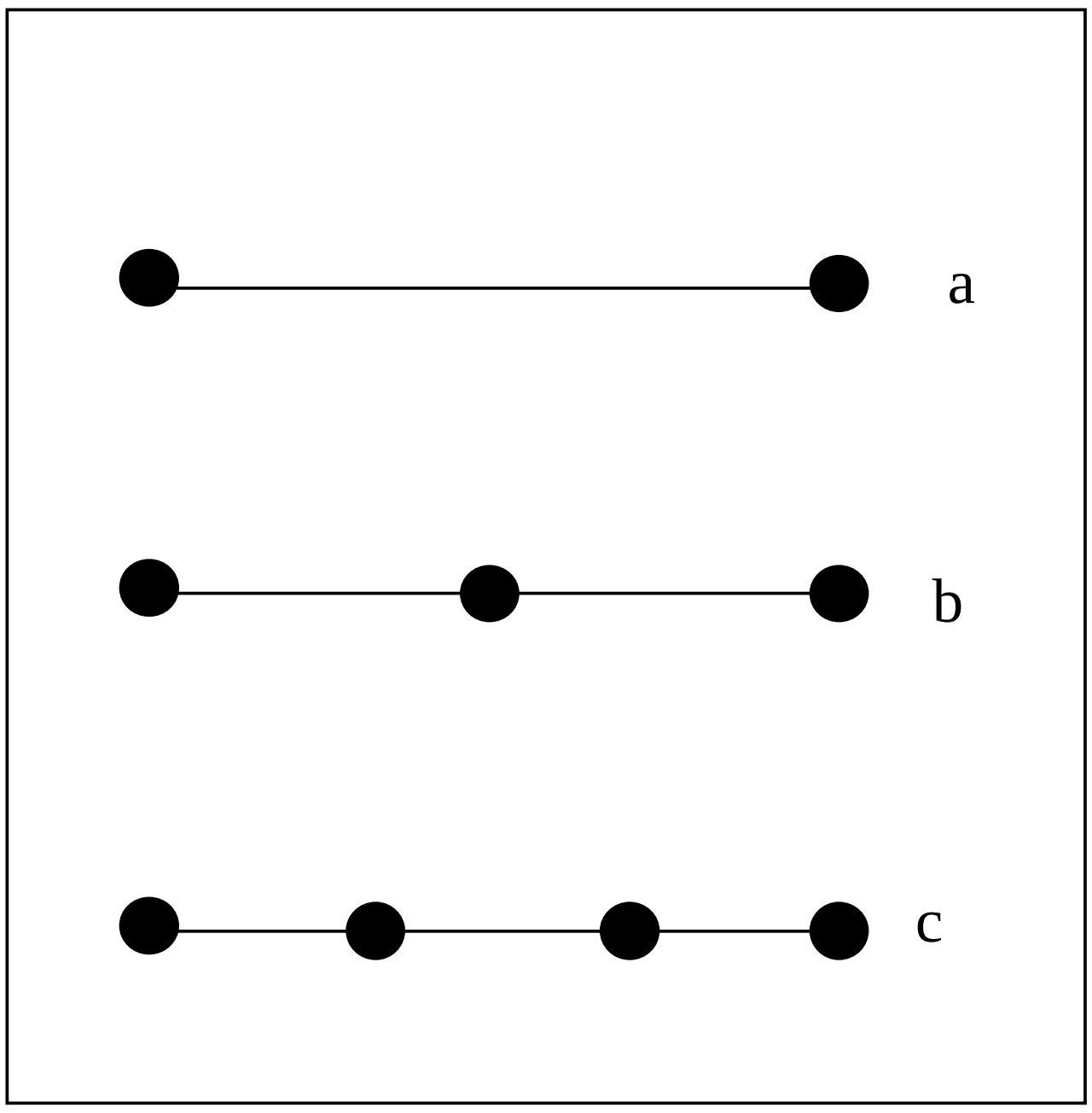

**Fig. 7**

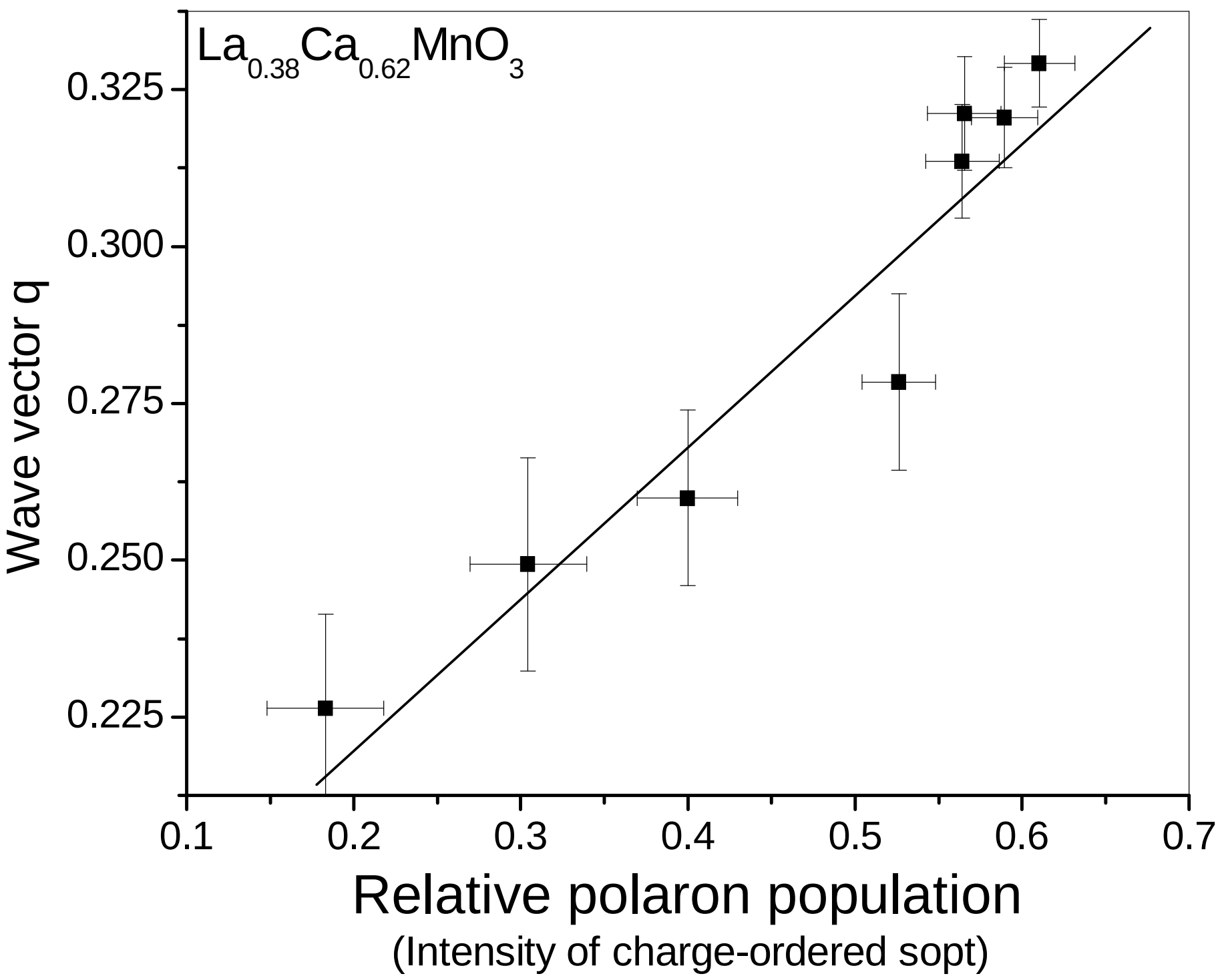

**Fig. 8**